\documentclass[conference]{IEEEtran}

\usepackage{graphicx,color}
\usepackage{amsthm}
\usepackage{amssymb}
\usepackage{amsmath}

\newtheorem{definitionenv}{Definition}
\newtheorem{lemmaenv}[definitionenv]{Lemma}
\newtheorem{theoremenv}[definitionenv]{Theorem}
\newtheorem{corollaryenv}[definitionenv]{Corollary}
\newtheorem{propositionenv}[definitionenv]{Proposition}
\newtheorem{conjectureenv}[definitionenv]{Conjecture}
\newtheorem{exampleenv}[definitionenv]{Example}
\newtheorem{app-lemmaenv}[section]{Lemma}

\newenvironment{lemma}{\begin{lemmaenv}\rm}{\end{lemmaenv}}
\newenvironment{theorem}{\begin{theoremenv}\rm}{\end{theoremenv}}
\newenvironment{corollary}{\begin{corollaryenv}\rm}{\end{corollaryenv}}
\newenvironment{example}{\begin{exampleenv}\rm}{\end{exampleenv}}

\newenvironment{conjecture}{\begin{conjectureenv}\rm}{\end{conjectureenv}}
\newenvironment{app-lemma}{\begin{app-lemmaenv}\rm}{\end{app-lemmaenv}}

\newcommand{\ba}{{\bf a}}
\newcommand{\bb}{{\bf b}}

\newcommand{\e}{{\bf e}}

\newcommand{\bg}{{\bf g}}

\newcommand{\bs}{{\bf s}}

\newcommand{\bu}{{\bf u}}
\newcommand{\bv}{{\bf v}}
\newcommand{\bw}{{\bf w}}
\newcommand{\x}{{\bf x}} 
\newcommand{\by}{{\bf y}}
\newcommand{\bz}{{\bf z}}

\newcommand{\cA}{{\cal A}}

\newcommand{\cN}{{\cal N}}

\newcommand{\cS}{{\cal S}}

\newcommand{\mF}{{\mathbb F}}

\newcommand{\be}{\begin{equation}}
\newcommand{\ee}{\end{equation}}

\newcommand{\bea}{\begin{eqnarray*}}
\newcommand{\eea}{\end{eqnarray*}}

\newcommand\wt{\mbox{{\rm wt}\,}}
\newcommand\Tr{\mbox{{\rm Tr}\,}}

\newcommand{\remove}[1]{}

\renewcommand{\le}{\leqslant}\renewcommand{\ge}{\geqslant}

\newcommand{\ds}{_{DS}}
\newcommand{\dsp}{_{DS}^\perp}

\begin{document}
\title{Correction of Data and Syndrome Errors by Stabilizer Codes}

\author{\IEEEauthorblockN{Alexei Ashikhmin}
\IEEEauthorblockA{Bell Laboratories\\
Alcatel-Lucent, 600 Mountain Ave\\
Murray Hill, NJ 07974\\
aea@reseach.bell-labs.com} \and \IEEEauthorblockN{Ching-Yi Lai}
\IEEEauthorblockA{
Institute of Information Science\\
Academia Sinica\\
No 128, Academia Road, Section 2\\
Nankang, Taipei 11529, Taiwan\\
cylai0616@iis.sinica.edu.tw}\and
\IEEEauthorblockN{Todd A. Brun} \IEEEauthorblockA{Communication
Sciences Institute\\
 University of Southern California\\
  Los Angeles, California, USA\\
  tbrun@usc.edu}}


\maketitle


\thispagestyle{empty}

\begin{abstract}
Performing active quantum error correction to protect fragile quantum states highly depends on the correctness of error information--error syndromes.
To obtain reliable error syndromes using imperfect physical circuits,
we propose the idea of quantum data-syndrome (DS) codes that are capable of correcting both data qubits and syndrome bits errors.
We study fundamental properties of quantum DS codes and provide several CSS-type code constructions of quantum DS codes.

\end{abstract}

\section{Introduction}
Quantum error-correcting codes provide a method of actively protecting quantum information \cite{LB13}.
To perform quantum error correction, we have to learn knowledge of errors, the \emph{error syndromes}, through quantum measurements.
However, the physical gates to do these operations are inevitably faulty, which may lead to incorrect data recovery.

In \cite{ALB14}, the authors introduced the so-called syndrome-measurement (SM) codes to perform robust syndrome measurement of stabilizer codes with Shor syndrome extraction \cite{DS96}.
Herein we consider a more general scheme of quantum stabilizer codes that are capable of correcting data qubit errors and syndrome bit errors simultaneously.
These codes are called {\em quantum data-syndrome codes} .
This idea is also independently studied by Fujiwara in \cite{Fuji14}.

We first define the quantum DS codes and related notions of minimum distance and split weight enumerators. Using
generalized MacWilliams  identities, we derive Singleton and Hamming type upper bounds on the code size of degenerate quantum DS codes, generalizing  the approach proposed in \cite{AL99}.
Next we study the properties of random quantum DS codes. Finally, we consider the CSS-type quantum DS codes and provide DS code constructions from CSS-type quantum cyclic codes and quantum LDPC codes.


\section{Quantum Data-Syndrome Codes}
Let $Q$ be an  $[[n,k]]$ stabilizer code defined by a stabilizer group $\mathcal{S}$ with
 (Pauli) generators  $g_1,g_2,\cdots,g_{n-k}$ and corresponding vectors $\bg_1,\ldots,\bg_{n-k}\in \mF_4^n$.
Denote
$$
H=\left[\begin{array}{c}
\bg_1\\
\vdots\\
\bg_{n-k}\end{array}\right].
$$
(Throughout the text for a Pauli operator $a$ acting on $n$ qubits  we denote by $\ba\in \mF_4^{n}$ the corresponding vector, see \cite{CRSS98} for details.)
Let $C$ be the classical $[n,n-k]$ code over $\mF_4$ generated  by the rows of $H$, and $C^\perp$ be its dual with respect
to the trace inner product:
$$
 \x*\by=\Tr^{\mF_4}_{\mF_2}(\sum_{i=0}^n x_i\overline{y}_i), \x,\by\in \mF_4^n,
 $$
where $\overline{y}_i$ denotes conjugation in $\mF_4=\{0,1,\omega,{\omega^2}\}$ with $\bar{0}=0,\bar{1}=1,
\bar{\omega}=\omega^2,$ and $\bar{\omega^2}=\omega$.
 We have $C\subseteq C^{\perp}$ since $\bg_i *\bg_j=0$ for all $i,j$.
Suppose a quantum state $|\psi \rangle\in Q$
 is corrupted by an error (Pauli) operator $e$ and let $\e\in \mF_4^n$  be the corresponding vector. Then the syndrome of $e$ is
 $$
 \bs^T=H*\e,
 $$
 where $\bs\in \mF_2^{n-k}$.
 One can use the syndrome $\bs$ for finding the most probable error vector $\e'$. If $\e'\in C+\e$,
 then applying $e'$ to the corrupted state will reconstruct the original state $|\psi \rangle$.

 One of the most difficult problems in quantum error correction is, however, that the syndrome $\bs$ itself
 could be measured with an error. So instead of the true vector $\bs\in\mF_2^{n-k}$, we may get, after measurement, a vector
 $\widehat{\bs}=\bs+\bz,$ for $\bz\in \mF_2^{n-k}$.
 In the following, we discuss stabilizer codes
that  are capable of correcting  both data errors and syndrome errors.

 Let us denote
 \begin{equation}\label{eq:HDS}
 H_{DS}=[HI_{n-k}],
 \end{equation}
 where $I_{n-k}$ is considered as a matrix over $\mF_2$.
To shorten notation, we will use $m=n-k$.
 Let us define codes
 \begin{align*}
 C_{DS}&=\{\bw=\bu H_{DS}: \bu\in \mF_2^m\}\subset \mF_4^n\times\mF_2^m, \mbox{ and }\\
 C_{DS}^\perp&=\{\bv: \bw \star \bv=0, \forall \bw\in C_{DS}\}\subset \mF_4^n\times\mF_2^m,
 \end{align*}
where  $C_{DS}^\perp$ is the dual code of $C_{DS}$ with respect to the inner product:
 $$
 \x\star\by=\Tr^{\mF_4}_{\mF_2}(\sum_{i=0}^n x_i\overline{y}_i)+\sum_{j=0}^m x_{n+j}y_{n+j},
 $$
  for $ \x,\by\in \mF_4^n\times \mF_2^m.$
Note that for any value of $k$ we have
\begin{equation}\label{eq:|Cperp|=2^2n}
|C^\perp_{DS}|=2^{2n}.
\end{equation}

 We will say that a stabilizer code $Q$ together with a particular choice of generators in (\ref{eq:HDS}) is a
 {\em quantum data-syndrome code} $Q\ds$ and that $C\ds^\perp$ is a
 {\em data-syndrome} code. We will say $C\ds^\perp$ (or $Q\ds$) has {\em quantum length} $n$, {\em quantum dimension} $k$, and {\em quantum size} $2^k$.

 Fujiwara, \cite{Fuji14}, noticed that error-correcting capabilities of $C\ds^\perp$ depend on the choice of generators in~(\ref{eq:HDS}).
In consequence of a bad choice of generators, even a single syndrome error may be uncorrectable.
On the contrary, choosing generators properly, we can get a code capable of correcting simultaneously  multiple
data and syndrome errors.

 In \cite{ALB14}, in order to correct syndrome errors, the authors proposed to get additional, say $r$, syndrome bits
 by measuring generators (not necessary distinct) $g_1',\ldots,g_r'\in \cS$. Let
 $$
H'=\left[\begin{array}{c}
\bg_1'\\
\vdots\\
\bg_r'\end{array}\right].
$$
 Then the approach from \cite{ALB14} is equivalent to considering the code $C\dsp$ with the parity check matrix
 $$
 H\ds'=\left[\begin{array}{ccc}
 H & I_{n-k} & 0 \\
H' & 0 & I_r
\end{array}\right].
$$
This parity check matrix can be transformed into the form
$$
H\ds''=\left[\begin{array}{ccc}
 H & I_{n-k} & 0 \\
0 & A & I_r
\end{array}\right],
$$
where $A$ is a binary matrix.
Now decoding can be done by first  using any decoder of the code with the parity check matrix {$[A\ I_{r}]$}
to correct syndrome errors, followed by the decoding  with respect to $H\ds$.
We will first concentrate on the scenario when only $n-k$ syndrome bits are measured, and then consider code constructions with additional syndromes.

\section{Mimimum Distance and Split Weight Enumerators}
Below we discuss the error correction capabilities of DS codes.
Consider  $\e\ds=(\bg,{\bf 0}),$ for $\bg \in C$.
 Since $\bg\in C$ we have $g\in \cS$ and thus $\e\ds$ is harmless.
 If $\e\ds=(\e,\bz)\in C\dsp\setminus \{(\bg,{\bf 0}): \bg\in C\}$, then $\e\not\in C$. Therefore the operator $e$ does not belong to $\cS$ and
 acts on $Q$ nontrivially. Since $H_{DS}\star \e\ds={\bf 0}^T$ by definition, we conclude that such $\e\ds$ is an undetectable and
 harmful error.
Naturally, the weight $\wt(\e,\bz)$ is defined as the number of its nonzero entries.
We define the {\em minimum distance} $d$ of $Q\ds$ (equivalently $C^\perp_{DS}$) as the minimum weight of any element in
 $$
C\dsp\setminus \{(\bg,{\bf 0}): \bg\in C\}.
 $$
It is not difficult to see that $Q\ds$ (or equivalently $C\dsp$) can correct any error  $\e_{DS}=(\e,\bz)$ (here we do not assume $\e\ds\in C\dsp$) with $t_D=\wt(\e),t_S=\wt(\bz)$ if
$$
t_D+t_S<{d\over 2}.
$$
%

  Define the split weight enumerators of $C_{DS}$ and $C_{DS}^\perp$ by
 \begin{align*}
B_{i,j}= B_{i,j}(C_{DS})=|\{ & \bw\in C_{DS}: \wt(w_1,\ldots,w_n)=i,\\
 &\wt(w_{n+1},\ldots,w_{n+m})=j\}|,\\
B_{i,j}^\perp= B_{i,j}(C_{DS}^\perp)=|\{ & \bw\in C_{DS}^\perp: \wt(w_1,\ldots,w_n)=i,\\
 &\wt(w_{n+1},\ldots,w_{n+m})=j\}|,
 \end{align*}
 respectively.
 The minimum distance $d$ of $Q\ds$ implies
\begin{align}
B_{i,0}^{\perp}= \sum_{j=0}^{m} B_{i,j}, \text{ for } i=1, \cdots, d-1. \label{eq:stabilizer_number}
\end{align}
We also have
 \begin{align}
 B^\perp_{i,0}\ge &\sum_{j=1}^m B_{i,j}, i=d,\ldots,n, \mbox{ and } \label{eq:Bperp>B} \\
 B_{0,0}=B_{0,0}^\perp=1 \mbox{ and } &B_{i,0}=0, i\ge 1.\label{eq:Bi0=0}
\end{align}
We will say that $Q\ds$ is a {\em degenerate} quantum DS code if there exists $B_{i,j}>0$ for $i<d$.
Otherwise this is a {\em nondegenerate} quantum DS code.
For $0\le r\le m$, let us define $d(r)$ as the smallest integer such that
$$
 B_{d(0),0}^\perp>\sum_{j=1}^m B_{d(0),j}, \mbox{ and } B_{d(r),r}^\perp>0, r=1,\ldots,m.
$$
Then the minimum distance of $Q\ds$ is
 $$
 d=\min_{0\le r\le m} d(r)+r.
 $$

Denote the $q$-ary Krawtchouk polynomial of degree $i$ by
\begin{equation}\label{eq:Kraw}
K_i(x;n,q)=\sum_{j=0}^i (-1)^j (q-1)^{i-j}{x\choose j}{n-x\choose i-j}.
\end{equation}
Properties of Krawtchouk polynomials can be found in~\cite{AL99}.
Let $f(x,y)$ be a two variable polynomial and its maximal degrees of $x$ and $y$ be $d_x\le n$ and $d_y\le m$, respectively. Then the following Krawtchouk expansion of this polynomial holds:
\begin{equation}\label{eq:KrawExpension}
f(x,y)=\sum_{i=0}^{d_x} \sum_{j=0}^{d_y} f_{i,j} K_i(x;n,q_1) K_j(y;m,q_2),
\end{equation}
where
$$
f_{i,j}={1\over q_1^n q_2^m} \sum_{x=0}^n \sum_{y=0}^m f(x,y)K_x(i;n,q_1)K_y(j;m,q_2).
$$

Using pretty much standard arguments \cite{MS77}, we get the following generalization of the MacWilliams identities.
\begin{theorem}
\begin{equation}\label{eq:MacW}
B_{l,r}={1\over {4^n}}\sum_{i=0}^n\sum_{j=0}^m B_{i,j}^\perp K_l(i;n,4) K_r(j;m,2).
\end{equation}
\end{theorem}
Like \cite{CRSS98,LBW13}, for small $n$ one could apply linear programming techniques to find bounds for quantum DS codes  
\begin{example}
Consider $n=7, m=6, d=3$. 
From MAPLE, both the primal and dual liner programs have solutions.
This means that a $[[7,1,3]]$ code may be capable of fighting a syndrome bit error by measuring only six stabilizer generators.
Indeed, it is the case, as  shown by Fujiwara~\cite{Fuji14}.
\end{example}
\section{Upper Bounds on Unrestricted (Degenerate and Non-Degenerate) DS codes}
We generalize the  approach suggested in \cite{AL99}.
Let $1\le d_D\le n$
be an integer and
$$
\cN=\{(i,j): 0\le i\le n, 1\le j\le m\}.
$$
Let also $\cA\subset \cN$ and $\overline{\cA}=\cN\setminus \cA$. We would like to upper bound {\em quantum code rate} $R=k/n$  under the conditions:
\begin{align}
B_{i,0}^\perp&=\sum_{j=0}^m B_{i,j}, i=0,\ldots,d_D-1, \label{bound_cond1}\\
B_{i, j}^\perp&=0, (i,j)\in \cA.\label{bound_cond2}
\end{align}
Let $f(l,r)$ be an {\em arbitrary} polynomial with nonnegative coefficients $f_{i,j}$ and satisfying the conditions:
\begin{align}
f(l,0)\le 0, &\mbox{ if } l\ge d_D, \mbox{ and} \label{eq:polin_cond1}\\
f(l,r)\le 0, &\mbox{ if } (l,r) \in \overline{\cA}\label{eq:polin_cond2}.
\end{align}
\begin{theorem}\label{thm:gen_upper_bound}
\begin{enumerate}

\item For non-degenerate $C^\perp_{DS}$ it must hold
\be\label{eq:non-degBound}
f(0,0)/ f_{0,0}\ge 2^{2n}.
\ee
\item For unrestricted $C^\perp_{DS}$ it must hold
\begin{equation}\label{eq:general_bound}
\hspace{-0.5cm} \max\left\{ {f(0,0)\over f_{0,0}}, \max_{1\le l\le d_D-1} {f(l,0)\over \min_{1\leq j\leq m} f_{l,j}}\right\} \ge 2^{2n}.
\end{equation}
\end{enumerate}
\end{theorem}
\proof
We prove the second claim.
Denote $M=|C_{DS}^\perp|=2^{2n}$. Using (\ref{eq:MacW}), (\ref{eq:KrawExpension}), and (\ref{eq:Bperp>B}), and we get
\begin{align*}
&M\sum_{i=0}^{d_D-1} \sum_{j=0}^m f_{i,j}B_{i,j}\le M \sum_{i=0}^n\sum_{j=0}^m f_{i,j}B_{i,j}\\
=&M \sum_{i=0}^n\sum_{j=0}^m f_{i,j}{1\over M} \sum_{l=0}^n\sum_{r=0}^m B_{l,r}^\perp K_i(l;n,4)K_j(r;m,2)\\
=&\sum_{l=0}^{n} B_{l,0}^\perp f(l,0)+\sum_{(i,j)\in \cA} B_{i,j}^\perp f(i,j)+\sum_{(i,j)\in \overline{\cA}}
B_{i,j}^\perp f(i,j)\\
\le & \sum_{l=0}^{d_D-1} B_{l,0}^\perp f(l,0)=\sum_{l=0}^{d_D-1} \sum_{j=0}^m B_{l,j} f(l,0)
\end{align*}
From this, using (\ref{eq:Bi0=0}), we get
\begin{align*}
2^{2n}\le &{\sum_{l=0}^{d_D-1}\sum_{j=0}^m B_{l,j} f(l,0) \over \sum_{i=0}^{d_D-1}\sum_{j=0}^m f_{i,j}B_{i,j}}\\
 \le &\max\left\{ {f(0,0)\over f_{0,0}}, \max_{1\le l\le d_D-1} {f(l,0)\over \min_{1\leq j\leq m} f_{l,j}}\right\}
\end{align*}
\QED


\subsection{Singleton Bound}\label{sec:SinB}
For getting a bound on DS codes with minimum distance $d$ it is enough to choose
\be\label{eq:cA for codes with min dist d}
d_D=d \mbox{ and } \cA=\{(i,j): j\geq 1,  0\leq i+j\le d-1\}.
\ee
Let
$$f(x,y)=\frac{4^{n-d+1}2^m}{{n\choose d-1}} {n-x \choose n-d+1}{m-y\choose m}.$$
It can be seen that $f(x,y)=0$ if $x\geq d$ or $y\geq 1$. It is easy to obtain that
 $f_{x,y} = {{n-x\choose d-1}/{n\choose d-1}}\geq 0.$
Thus $f(x,y)$ satisfies constraints (\ref{eq:polin_cond1}) and (\ref{eq:polin_cond2}).
After simple computations we get
$$
\max\left\{f(0,0), \max_{1\leq i\leq d-1} \frac{f(i,0)}{\min_{1\leq j\leq m}f_{i,j}}   \right\}=4^{n-d+1}2^m.
$$
This and Theorem \ref{thm:gen_upper_bound} lead to the following result
\begin{theorem} ({\bf Singleton Bound}) For an unrestricted (non-degenerate or degenerate)  DS code we have
$$
k\leq n-2(d-1).
$$
\end{theorem}
Thus we proved that the Singleton bound for quantum stabilizer codes also holds for unrestricted quantum DS codes.


\subsection{Hamming Bound}
Let $C^\perp_{DS}$ be a non-degenerate DS code with minimum distance $d=2t+1$.
Using standard combinatorial arguments, we get that $k\le \tilde{k}$, where
$\tilde{k}$ is the largest integer such that
\begin{align}\label{eq:HBnondeg}
 2^{2n} & \leq 4^n2^{n-\tilde{k}}/\sum_{i=0}^{t} {n\choose i}3^i \sum_{j=0}^{t-i} {n-\tilde{k}\choose j}.
\end{align}
This is the {\em Hamming Bound  for non-degenerate} DS codes.

Let $d_D$ and $\cA$ be defined as in (\ref{eq:cA for codes with min dist d}). Let $f^{(k)}(l,r)$ be the polynomial defined by the coefficients
$$
f^{(k)}_{i,j}=\left(\sum_{h=0}^{t}  K_h(i;n,4) \sum_{s=0}^{(t-h)/\lambda}K_s(j;m,2)\right)^2.
$$
Here $\lambda$ is a parameter over which we will optimize our bound.
After some computations, we obtain
$$
f^{(k)}(l,r)=4^n2^m \sum_{i=0}^{t}\sum_{j=0}^{t} \beta(r,i,j) \sum_{u=0}^{t-\lambda} i\sum_{v=0}^{t-\lambda} j\sum_{h=0}^{n-l} \alpha(l,u,v,h),
$$
where
\begin{align}
\alpha(l,i,j,h)=&{l\choose 2l+2h-i-j}{n-l\choose h}{2l+2h-i-j \choose l+h-j}
\nonumber \\
&2^{i+j-2h-l}3^h, \label{eq:alpha}
\mbox{ and }\\
\beta(r,u,v)=&{m-r\choose (u+v-r)/2}{r\choose (u-v+r)/2}. \label{eq:beta}
\end{align}
This leads to the following result.
\begin{theorem}\label{f(x,y)=0}
For $x+y\ge d$ we have $f^{(k)}(x,y)=0$.
\end{theorem}
Thus $f^{(k)}(x,y)$ satisfies constraints (\ref{eq:polin_cond1}) and (\ref{eq:polin_cond2}).

%

\begin{figure}[h]
\vspace{-0.5cm}
\[
\includegraphics[width=6.5cm]{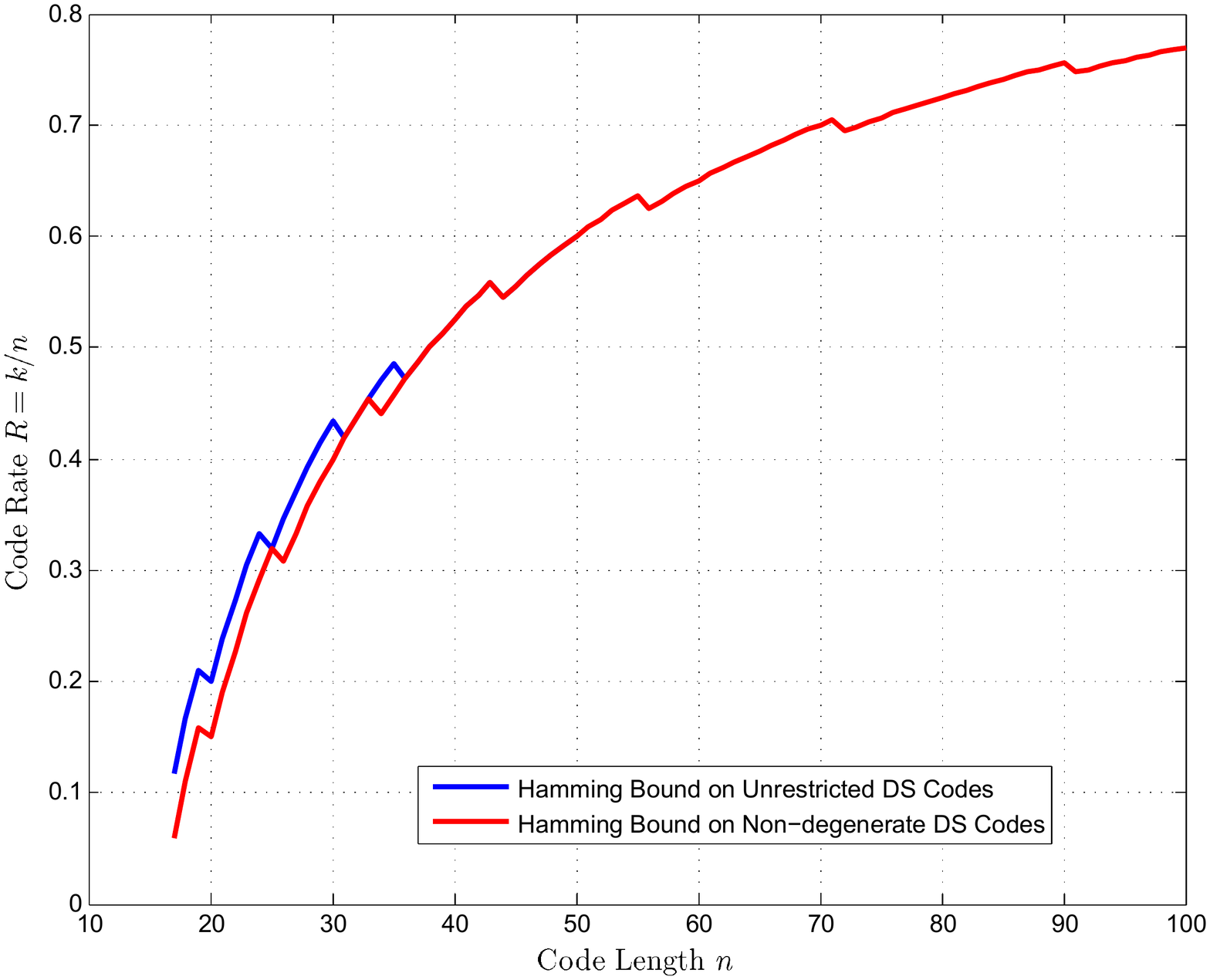}
\]
\vspace{-1.0cm}
\caption{Hamming Bounds for non-degenerate and unrestricted DS codes, $d=7$.
 } \label{fig:Hamming bound} 
\end{figure}

\begin{theorem}({\bf Hamming bound for unrestricted DS codes}.) For an unrestricted DS code we have
$k\le \overline{k}$, where
 $\overline{k}$ is the largest integer such that
\begin{align}\label{eq:HBdeg}
&\min_{1\le \lambda \le t+1} \max\left\{ {f^{(\overline{k})}(0,0)\over f^{(\overline{k})}_{0,0}}, \max_{1\le l\le d_D-1} {f^{(\overline{k})}(l,0)\over \min_{1\leq j\leq m} f^{(\overline{k})}_{l,j}}\right\}\nonumber \\
&\ge 2^{2n}.
\end{align}
\end{theorem}
For $d=7$ Hamming bounds (\ref{eq:HBnondeg}) and (\ref{eq:HBdeg}) are shown in Fig.\ref{fig:Hamming bound}. For small values of $n$
the bound for unrestricted DS codes is only marginally weaker than (\ref{eq:HBnondeg}) and for $n\ge 36$ these bounds coincide.
We observed the same behavior for other values of $d$. So we make the conjecture.
\begin{conjecture} For any $d$ there exists $n(d)$ such that for $n\ge n(d)$ the Hamming bound (\ref{eq:HBnondeg}) holds for unrestricted
DS codes.
\end{conjecture}

In \cite{Fuji14} Fujiwara obtained the {\em hybrid Hamming bound} for non-degenerate DS codes that can correct any
 $t_D$ data and $t_S$ syndrome errors:  $k\le \hat{k}$, where $\hat{k}$ is the largest integer s.t.
\begin{equation}\label{eq:HybridHamBound}
 2^{2n}\le 2^{2n}2^{n-\hat{k}}/\sum_{i=0}^{t_D}\sum_{j=0}^{t_S} {n \choose i}3^i {n-\hat{k}\choose j}.
 \end{equation}
For obtaining hybrid bounds using Theorem \ref{thm:gen_upper_bound}, we choose
$$
d_D=2t_D+1 \mbox{ and } \cA=\{ (i,j): 0\le i\le 2t_D \mbox{ and } 1\le j\le 2t_S\}.
$$
Let
$
f^{(k)}(l,r)=4^n2^m \sum_{i=0}^{t_D}\sum_{j=0}^{t_D} \sum_{h=0}^{n-l} \alpha(l,i,j,h) \times \sum_{u=0}^{t_S}\sum_{v=0}^{t_S}
\beta(r,u,v).
$
\begin{theorem} For this polynomial we have $f^{(k)}_{i,j}\ge 0$, $f^{(k)}(l,r)=0$ if $(l,r)\in \overline{\cA}$,
and that $f^{(\hat{k})}(0,0)/f^{(\hat{k})}_{0,0}$ is equal to the right hand side of (\ref{eq:HybridHamBound}).
\end{theorem}
So we got a different proof of (\ref{eq:HybridHamBound}). We can not use this polynomial
for unrestricted DS codes, since for some $i\le d_D-1$ we have  $f^{(k)}_{i,j}=0$. Finding good polynomials for
hybrid bounds on unrestricted DS codes is an open problem.

\section{Random DS Codes}
The enumerators $B_{i,j}^\perp$ define the decoding error probability of a DS code in a number of communication/computational scenarios. Below we study the behavior of   $B_{i,j}$ and $B_{i,j}^\perp$ of random DS codes.
In particular, for a given $r/n$ we are interested in the behavior of $d(r)/n$ when $n\rightarrow\infty$.

Let $\cS_{n,k}$ be the set of all codes with the generator matrix of form (\ref{eq:HDS}), and $\cS_{n,k}^\perp$ be the set of the codes dual to codes from $\cS_{n,k}$.
Define the average weight distributions of codes from $\cS_{n,k}$ and $\cS_{n,k}^\perp$ by
\begin{align*}
\overline{B}_{i,j}&={1\over |\cS_{n,k}|} \sum_{C\in \cS_{n,k}} B_{i,j}(C),\\
\overline{B}_{i,j}^\perp&={1\over |\cS_{n,k}^\perp|} \sum_{C^\perp\in \cS_{n,k}^\perp} B_{i,j}(C^\perp).
\end{align*}
To find $\overline{B}_{i,j}$ and $\overline{B}_{i,j}^\perp$ we need the following lemma.
\begin{lemma}
$$
|\cS_{n,k}|=|\cS_{n,k}^\perp|=\prod_{r=0}^{n-k-1} {(2^{2(n-r)}-1)(2^{n-k}-2^r)\over 2^{r+1}-1}.
$$
Any vector $\bw=(\ba,\bb),\ba\in \mF_4^n\setminus {\bf 0},\bb\in \mF_2^{n-k}\setminus {\bf 0}$ is contained
in
$$
{2^{n-k}-1\over 4^n-1}\prod_{r=0}^{n-k-1} {(2^{2k}-1)\over 2^{r+1}-1}
$$
codes from $\cS_{n,k}$.
\end{lemma}
Using this Lemma, we obtain the following result.
\begin{theorem}
\begin{align*}
\overline{B}_{0,0}&=\overline{B}_{0,0}^\perp=1,\\
\overline{B}_{i,j}&={1\over 4^n-1}{n\choose i}3^i {n\choose j},i>0 \mbox{ or }j>0,\\
\overline{B}_{i,j}^\perp&={4^n\over (4^n-1)2^k}{n\choose i}3^i {n\choose j}, i,j>0,\\
\overline{B}_{i,0}^\perp&={2^{2n-k}-1\over 4^n-1}{n\choose i}3^i, i>0\\
\overline{B}_{0,j}^\perp&=0, j>0.
\end{align*}
\end{theorem}

Using Markov's inequality and the union bound we get:
$$
\Pr(B_{i,j}(C^\perp)\le (nm)^{1+\epsilon} \overline{B}_{i,j}^\perp \mbox{ for all } i,j)\ge 1-{1\over (nm)^{\epsilon}},
$$
for any $\epsilon>0$. Hence there exists a code $C^\perp\in {\cS}_{n,k}^\perp$ such that
\begin{equation}\label{eq:Bij_expur}
B_{i,j}(C^\perp)\le (nm)^{1+\epsilon} \overline{B}_{i,j}^\perp, \forall i,j.
\end{equation}
Let $H(x)=-x\log_2(x)-(1-x)\log_2(1-x)$ be the binary entropy function. Denote
$$
R=k/n, \omega=i/n \mbox{ and } \gamma=j/n.
$$
With these notations, for codes satisfying (\ref{eq:Bij_expur}), we have
\begin{align}
b_{\omega,\gamma}^\perp&={1\over n}\log_2 B_{\lfloor \omega n\rfloor,\lfloor \gamma n\rfloor}^\perp \label{eq:splitGV}\\
&=H(\omega)+\omega \log_2(3)+RH(\gamma R)-R+o(1).\nonumber
\end{align}
Since $B_{i,j}(C^\perp)$ are integers, for sufficiently large $n$ we get that  $B_{\lfloor \omega n\rfloor,\lfloor \gamma n\rfloor}^\perp=0$ as soon as $b_{\omega,\gamma}^\perp<0$. Summarizing this we get the following result.
\begin{theorem}
For a given $\gamma$ let $\omega^*$ be the root of
$$
H(\omega)+\omega \log_2(3)+RH(\gamma R)-R=0.
$$
Then there exist DS codes with
$
B_{\lfloor \omega n\rfloor,\lfloor \gamma n\rfloor}^\perp=0, \omega<\omega^*.
$
\end{theorem}
This Theorem can be used for getting an estimate on the decoding fidelity of a random DS code. This estimate will be obtained in another work.

\section{CSS-type Quantum DS Codes}
In the following we focus on CSS-type DS codes. Suppose we have an $[[n,k,d]]$ CSS code defined by a binary parity-check matrix $H$. 
Let
\begin{align}
H\ds=\left(\begin{array}{cc|cc}{H}& {I}_{m/2}  & 0 & 0\\0 & 0 &{H} &{I}_{m/2} \end{array}\right). \label{eq:modified check matrix}
\end{align}
Suppose $\tilde{H}=[H {I_{m/2}}]$ is a parity-check matrix of a classical $[n',k',d']$ code.
Then the minimum distance of the  corresponding DS code is $d'\leq d$ and we call this an $[[n,k,d',m ]]$ quantum DS code,
where $m$ represents the number of generators to be measured.
Next we discuss how to extend a given parity-check matrix $H$ so that $\tilde{H}$ has minimum distance $d$.

An observation is that if no columns of $H'$ are of weight one, the distance of $\tilde{H}$ is at least $3$.
When $d>3$, it becomes nontrivial. 
Another observation is that the syndromes are the column space of $H'$.
We may treat $H'^T$ as a generator matrix of a classical code and the syndrome bit errors are normal bit flip errors.
As a consequence, if the column space of $H'$ has minimum distance no less than $d$, then these syndromes are robust up to $\lfloor \frac{d-1}{2}\rfloor$ qubit and syndrome bit errors.

\begin{theorem} \label{thm:cyclic code}
Suppose $H$ is a parity-check matrix of an $[n,k,d]$ classical dual-containing  cyclic code.
Then there exists an $[[n,2k-n,d, m\leq 2n]]$ quantum DS code. 
\end{theorem}
%



Quadratic-residue (QR) related codes provide a family of  such quantum DS codes \cite{MS77,LLu07}.
%

\begin{theorem} \label{thm:QR_double_even}
For $p=8j-1$,  
there are $[[p,1,d]]$ doubly-even CSS codes with $d^2 - d + 1 \geq p$,
where CNOTs, Hadamards, and Phase gates can be done transversally.
Moreover, there are $[[p,1,d,m\leq 2p]]$ quantum DS codes.
\end{theorem}
%
%

\begin{example}
Consider the QR ocde with $p=23$.
Suppose $H'$ is cyclicly generated by $q(x)$ with  $s\geq 11$ rows.
As can be seen in Table \ref{tb:QR23}, a $[[23,1,7,40]]$ quantum DS code exists with only $40$ redundant stabilizers, instead of $46$.
\begin{table}[h]
\[
\begin{tabular}{|c|c|c|c|c|c|c|c|c|c|c|c|}
  \hline
  $s$ & 11 & 12 & 13 & 14 & 15 & 16 & 17 & 18 & 19 &20 \\
\hline
  $d$ & 3 & 4 &4 & 4 & 5 & 5 & 5 & 6 & 6 & 7 \\
  \hline
\end{tabular}
\]  \caption{The distance of $\tilde{H}$ corresponding to  different number of rows. } \label{tb:QR23}
\end{table}

\end{example}

The family of quantum QR codes in Theorem \ref{thm:QR_double_even}, which includes the Steane code and the quantum Golay code, are
important in the theory of fault-tolerant quantum computation that are shown to have error thresholds. 
Here we have shown that these codes also induce quantum DS codes.


Finally, it is natural to consider the quantum DS codes induced from LDPC codes since a parity-check matrix
usually has some redundant rows and  hence there is additional error-correcting ability on the measurement errors.

Consider a $(\gamma,\rho)$-regular LDPC code.
The parity check matrix has the property that
each column consists of $\gamma$ $1$'s and  the number of $1$'s in common between any two columns is no greater than 1.
When $\gamma> 1$, it means that any single qubit error must have a syndrome of weight $\gamma>1$.
Any two-qubit data error must have a syndrome of weight at least $2\gamma-1$.
\begin{corollary}
Suppose there is an LDPC code that is self-orthogonal.
If $\gamma>1$, the induced quantum DS code has distance at least $3$.
if $\gamma>2$, the induced quantum DS code has distance at least $5$.
\end{corollary}

\begin{figure}[h]
\[
\includegraphics[width=7.5cm]{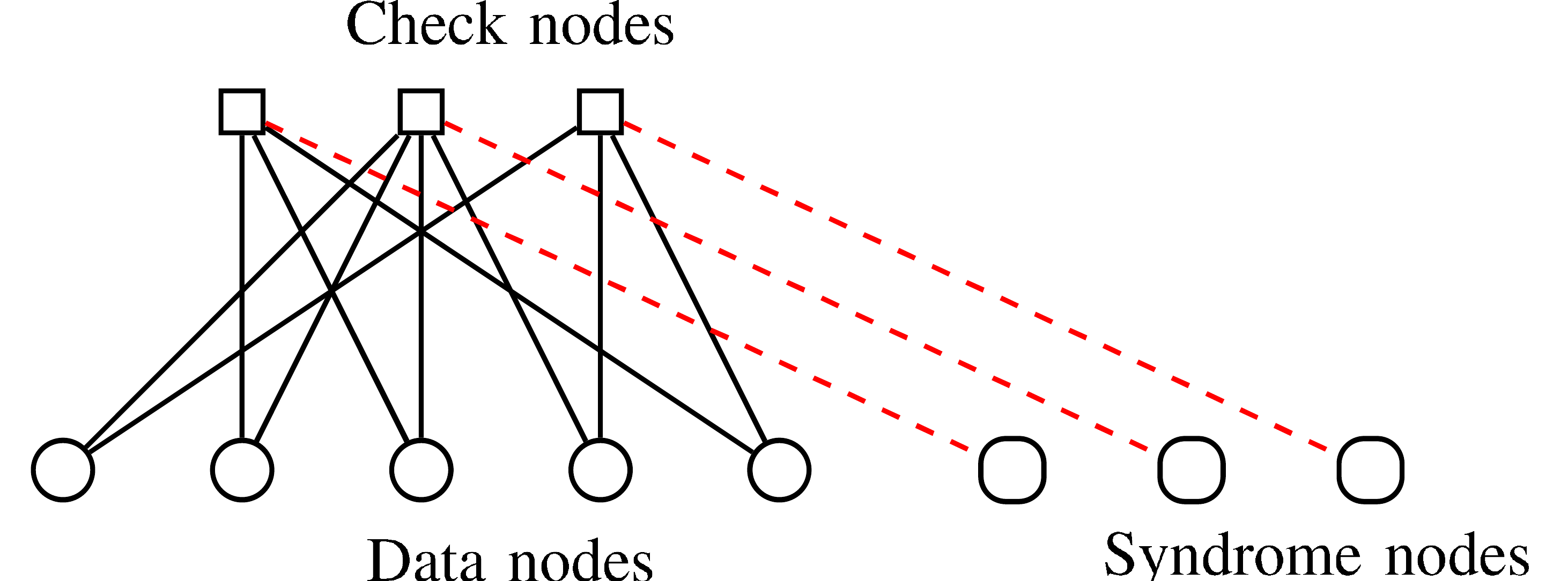}
\]
  \caption{
    A modified Tanner graph.
  }\label{fig:modified tanner graph}
\end{figure}

We may modify the Tanner graph for decoding as in Fig. \ref{fig:modified tanner graph}.
The bit nodes in the dashed box represent the syndrome bits.
In the (belief propagation) iterative decoding, we can assign distinct initial error rates for each bit:
the data qubit error rates are different from syndrome bit error rates; moreover,  the error rates for each syndrome bit are generally different.
Although LDPC codes may be inherently robust to some syndrome bit errors, we still can introduce more redundant check notes and corresponding syndrome nodes.

%
%
%
%


%
\section{Discussion}
If we use a large  syndrome measurement code, the errors in the quantum state may seriously accumulate in an error correction cycle.
Also, the redundant stabilizers may have high weight, which increases the probability of measurement errors.
Therefore, an SM code has to be carefully chosen.

When data qubit error rates and syndrome bit error rates are comparable, we may use a quantum DS code.
If error rates get higher, a quantum DS code with higher distance can be used.
However, there may not be many choices in some cases, such as quantum codes with a transversal $T$ gate.
Therefore, SM codes are preferred in this scenario.


\end{document}